\documentclass{optica-article}

\journal{opticajournal} 

\articletype{Research Article}

\usepackage{lineno}
\usepackage{wasysym}
\usepackage{hyperref}

\begin{document}

\title{Resonantly pumped tunable Tm,X:CaF\textsubscript{2} lasers: Effect of buffer ions (X = Y, La, Gd, and Lu)}

\author{Dominika Popelová\authormark{1,*}, Karel Veselský\authormark{1}, Pavel Loiko\authormark{2}, Abdelmjid Benayad\authormark{2}, Alain Braud\authormark{2}, Patrice Camy\authormark{2},\\Jan Šulc\authormark{1}, Helena Jelínková\authormark{1}}

\address{\authormark{1}Faculty of Nuclear Sciences and Physical Engineering, Czech Technical University in	Prague, Břehová 7, Prague, 11519, Czech Republic \\
\authormark{2}Centre de Recherche sur les Ions, les Matériaux et la Photonique (CIMAP), UMR 6252 CEA-CNRS-ENSICAEN, Université de Caen Normandie, 6 Boulevard Maréchal Juin, Caen Cedex 4, 14050, France\\}

\email{\authormark{*}dominika.popelova@fjfi.cvut.cz}

\begin{abstract*} 
We report on the eye-safe laser performance of Tm$^{3+}$-doped calcium fluoride crystals, modified with optically inactive ``buffer'' cations (Y$^{3+}$, Lu$^{3+}$, Gd$^{3+}$, and La$^{3+}$), under in-band diode-pumping at 1.68 \textmu m. In the free-running regime, the 1.5 at.\% Tm, 4 at.\% Y:CaF$_{2}$ laser operates with a high slope efficiency of 47\% with respect to absorbed pump power. By employing an MgF$_{2}$ birefringent plate, the emission wavelength of this laser is continuously tuned over 242~nm (1820 – 2062~nm) highlighting its potential for broadly tunable and ultrafast oscillators. The spectroscopic properties of Tm$^{3+}$ ions in CaF$_{2}$ crystals with and without buffer ions are also compared.

\end{abstract*}

\section{Introduction}
\label{sec1}

Tunable lasers have emerged as indispensable tools across a wide range of applications due to their ability to emit light over a continuously adjustable wavelength range. This flexibility allows to precisely select the emission wavelength with a narrow linewidth to match specific absorption features, optimize the system performance or facilitate multiplexed measurements. Tunable lasers are essential for spectroscopy due to their ability to selectively probe molecular transitions \cite{Du2019}, for atmospheric monitoring \cite{Hinkley1976}, as well as for biomedical imaging \cite{Yun2004}, optical communications \cite{Al-Taiy2014} and material processing \cite{Duarte2009}. In solid-state lasers, the wavelength tuning is usually enabled by a diffraction grating, an etalon or a birefringent filter. A birefringent filter (BRF) is typically made of anisotropic materials (e.g., SiO$_{2}$ or MgF$_{2}$) and has a form of a thin plate \cite{Wei2023}. The tuning is achieved by rotating the filter around an axis perpendicular to its surface. Proper operation requires the incoming radiation to be linearly polarized. This can be accomplished by using polarizers or by tilting the filter at the Brewster angle, which also helps reduce the losses in the resonator. The BRF exhibits wavelength-dependent transmission maxima, which can be shifted by its rotation. The tuning range is determined by the spacing between two consecutive maxima, known as the free spectral range (FSR), which is inversely proportional to the filter thickness \cite{Svelto2010,Kobtsev1992,Zhu1990}.

Tunable lasers emitting in the vicinity of 2~\textmu m gather attention due to a unique combination of their eye safety, atmospheric transmission and strong molecular absorption features. One of the primary advantages of 2-\textmu m lasers is that such radiation is absorbed more readily in the eye’s cornea reducing the risk of retinal damage and enabling applications in LIDAR and remote sensing. This spectral region also contains the strong absorption lines of several gases, such as H$_{2}$O, CO$_{2}$, CH$_{4}$ and NH$_{3}$, making these lasers effective for trace gas detection and environmental monitoring of atmospheric pollutants, greenhouse gases and industrial emissions \cite{McComb2010}. In the medical field, the high tissue absorption at 2~\textmu m facilitates precise tissue ablation with minimal tissue damage \cite{Bilici2010}. Nowadays, thulium and holmium lasers are mainly used to address the 2-\textmu m range.

The material wise choice of laser gain media for broadly tunable lasers comprises media with strong spectral line broadening, either induced by the electron-lattice (vibronic) interaction in the case of transition-metal ions (vibronic lasers), or by inhomogeneous line broadening in the case of rare-earth ions with their intrinsically narrow 4f – 4f electronic lines. Rare-earth-doped disordered laser gain media allow for continuous wavelength tuning over tens to hundreds of nanometers, covering several electronic lines between Stark sub-levels of the involved multiplets. Rare-earth doped glasses (including glass fibers) naturally appear as the first choice; however, their use is limited by low thermal conductivity and low laser damage thresholds. Single crystals (and polycrystalline transparent ceramics) with a structure and / or compositional disorder present an alternative, even though the local disorder inevitably leads to a drop in their thermal conductivity as compared to ordered crystals. So far, various crystals (ceramics) doped with Tm$^{3+}$ ions and exploiting the structural disorder (such as CaGdAlO$_4$  \cite{Wang2016,Pan2019} or CNGG-type garnets \cite{Pan2018}), or compositional disorder in solid-solution compositions (such as “mixed” sesquioxide ceramics \cite{Wang2018,Zhao2020}), have enabled broadband tuning of laser emission around 2~\textmu m.

Among the disordered laser gain media, cubic calcium fluoride (CaF$_{2}$, called fluorite in the mineral form) crystals stand apart. They readily form solid solutions with rare-earth ions, CaF$_{2}$ – RF$_{3}$, or Ca$_{1-x}$R$_{x}$F$_{2+x}$ (accounting for charge neutrality), with the solubility limit up to $\sim$40 mol\%, retaining their cubic structure. Remarkably, rare-earth ions in CaF$_{2}$ tend to reside in close proximity to each other, forming clusters composed of R$^{3+}$ - F$_{i-}$ pairs (i – interstitial anion), whose geometry depends on the doping level but also on the nature of the rare-earth cation. Such crystals were called “anti-glasses” \cite{Kuznetsov2006} as they lack the short-range order while maintaining the long-range one (``islands of disorder in the sea of order'', a situation opposite to that in amorphous glasses). Several distinct cluster types may coexist and within each type (called a ``quasi-center''), further diversity of species may be revealed by low temperature spectroscopy. This strongly multi-site behavior results in smooth, structureless and broad absorption and emission bands. Remarkably, the attribution of electronic transitions is often not possible even at cryogenic temperatures when the phonon assisted processes are suppressed.

By combining the broadband emission behavior of rare-earth ions in CaF$_{2}$ with the inherently appealing properties of this fluoride host (namely, high thermal conductivity, negative dn/dT coefficient, low phonon energy, and broadband transparency), these crystals are currently seen as excellent candidates for the replacement of laser glasses in high power / energy facilities operating around 1~\textmu m (based on Nd$^{3+}$ or Yb$^{3+}$ ions). There is also a rising interest in CaF$_{2}$ crystals for 2-\textmu m and 3-\textmu m lasers \cite{Normani2023}. Note that CaF$_{2}$ can also be fabricated in the form of transparent ceramics even though the technology is still far from being mature \cite{Sulc2014}.

Calcium fluoride crystals codoped with several rare-earth ions would contain clusters of complex composition which enhance energy-transfer between them, as in the case of Yb$^{3+}$,Tm$^{3+}$ \cite{Soulard2017} or Tm$^{3+}$,Ho$^{3+}$ \cite{Eremeev2023} ion pairs. However, if the codopant is optically inactive, it will play the role of a ``buffer'' effectively suppressing the crosstalk between the neighboring active ions, a property widely exploited to enable laser emission from Nd$^{3+}$-doped CaF$_{2}$ crystals  \cite{Normani2016,Meroni2022}. The “buffer” ions X$^{3+}$ could be Y$^{3+}$, La$^{3+}$, Gd$^{3+}$, Lu$^{3+}$, or their combination, and the effect is increased on enhancing the buffer/laser ion codoping ratio. In addition to suppressed energy-transfer processes and reduced luminescence quenching, the variation of the cluster composition changes the multi-ligand environment around the laser-active ions, contributing to an altered crystal-field, and, consequently, Stark splitting of their manifolds, as well as the probabilities of radiative transitions. Finally, the mass difference between the ``buffer'' and active ions increases the local disorder, which could result in further spectral broadening (although this effect seems to be weaker than the disorder induced by the ion clustering itself). CaF$_{2}$ crystals codoped with laser-active Tm$^{3+}$ ions, on one side, and various buffers, have been reported. Veselský et al. reported on a diode-pumped cryogenically cooled Tm,Lu:CaF$_{2}$  reaching slope efficiency of 66\% and tuning range of 250~nm from 1785 to 2035~nm \cite{Veselsky2023}. Liu et al. reported on a diode-pumped Tm,Y:CaF$_{2}$ laser delivering 0.45~W at 1.94~\textmu m with a slope efficiency of 21.5\% and achieved 190~nm wide tuning range (1850-2040~nm) \cite{Liu2017a}. Zhang et al. demonstrated power scaling of a Tm,La:CaF$_{2}$ laser up to 4.05~W at 1.92~\textmu m and achieved a similar tuning range, 1840-2032~nm (192~nm wide) \cite{Zhang2018}.

The reduced crosstalk between active ions in complex clusters containing ``buffer'' ions can negatively affect laser schemes relying on cross-relaxation (CR) to achieve higher pump quantum efficiency (e.g., the 2-\textmu m transition of Tm$^{3+}$) or energy-transfer upconversion (ETU) to avoid the bottleneck effect (e.g., the 2.8-\textmu m transition of Er$^{3+}$). In the case of Tm$^{3+}$, there exists an alternative pumping scheme directly to the upper laser level (the $^{3}$H$_{6}$~$\rightarrow$~$^{3}$F$_{4}$ transition in absorption), referred to as resonant or in-band pumping  \cite{Loiko2019}. Its advantages over conventional pumping around 0.8~\textmu m include lower quantum defect, reduced heat loading, high laser efficiency, independence from the CR process removing the limitations of the Tm$^{3+}$ doping level, and operation entirely within the eye-safe spectral region \cite{Kratochvil2024a}. The Tm$^{3+}$,X$^{3+}$-codoped CaF$_{2}$ crystals are appealing for this type of pumping as they benefit from reduced ETU from the upper laser level which usually contributes to increased laser thresholds. In-band pumping of Tm:CaF$_{2}$ crystals has been exploited: Thouroude et al. reported on 1.25~W at 1.88~\textmu m from a Tm:CaF$_{2}$ laser pumped by a 1.61~\textmu m Er-Yb-codoped fiber laser \cite{Thouroude2020}.

In the present work, we report a comparative spectroscopic and laser study of five Tm$^{3+}$-doped calcium fluorite crystals, both singly-doped and codoped with various optically inactive ``buffer'' cations X$^{3+}$ = Y$^{3+}$, La$^{3+}$, Gd$^{3+}$, and Lu$^{3+}$. We focus on their wavelength-tuning potential around 2~\textmu m under in-band pumping. As a result, we demonstrate continuous tuning up to 242~nm (1820--2062~nm) with the Tm,Y:CaF$_2$ laser, highlighting the potential of these disordered gain media for ultrabroadband coherent light sources.

\section{Crystal growth and spectroscopy}

\subsection{Crystal growth}
\label{subsec1}

Single crystals of Tm,X:CaF$_{2}$ were grown by the Bridgman–Stockbarger method using graphite crucibles ($\diameter$: 8~mm; height: 40~mm). The starting reagents were CaF$_{2}$ (Sigma-Aldrich, purity: 4N) and rare-earth fluorides, RF$_{3}$, R = Tm, Y, La, Gd, and Lu, produced from commercial R$_{2}$O$_{3}$ powders (Alfa Aesar, purity: 4N) using an excess of NH$_{4}$HF$_{2}$ solution at 250 $^\circ$C, following the reaction R$_{2}$O$_{3}$ + 6 NH$_{4}$HF$_{2}$ $\rightarrow$ 2 RF$_{3}$ + 6 NH$_{4}$F + 3 H$_{2}$O. The obtained dry cakes were annealed at 650 $^\circ$C under an Ar atmosphere to remove the residual NH$_{4}$F and moisture. To avoid oxygen contamination, the growth was performed under an inert (Ar + CF$_{4}$) atmosphere. The doping level was 1.5~at.\% in Tm$^{3+}$ and 4.0~at.\% in X$^{3+}$ cations (with respect to Ca$^{2+}$). The growth charge was placed into the crucible, which was heated slightly above the melting point ($\sim$1418~$^\circ$C), and the melt was homogenized for about 4~h. The growth was achieved by translating the crucible in a vertical temperature gradient of 30 – 40~$^\circ$C/cm at a pulling rate of 1 – 3~mm/h. After completion of the growth, the crystals were slowly cooled down to room temperature over a period of 48~h. No additional annealing was applied. The crystals were transparent and nearly colorless. No spectroscopic signatures of Tm$^{2+}$ ions were detected. The segregation coefficient for Tm$^{3+}$ ions was close to unity, and the doping gradient over the entire crystal boule did not exceed 0.2~at.\% (as evaluated by absorption measurements).

\begin{figure}[h]
	\centering
	\includegraphics[width=0.5\textwidth]{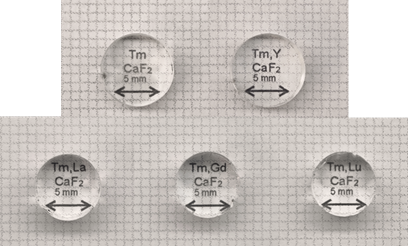}
	\caption{Photographs of the as-grown and laser grade polished 1.5 at. \% Tm$^{3+}$, 4 at. \% X$^{3+}$:CaF$_{2}$ (X = no buffer, Y, La, Gd, or Lu) active media.}
	\label{fig:Crystals}
\end{figure}

The laser elements were cut from the central part of the as-grown boules having a thickness of about 5~mm and a diameter of 7~mm. They were polished on both sides with good parallelism and left uncoated, as shown in Fig.~\ref{fig:Crystals}.

\subsection{Optical spectroscopy}

Calcium fluoride crystals suffer from relatively low transition cross-sections for the dopant ions (at moderate to high doping levels, where the ion clustering is dominant). This leads to low pump absorption efficiency and low gain in these materials. Physically, this is related to strong inhomogeneous spectral line broadening. Codoping with buffer ions helps to change the cluster composition and (in some cases) enhance the transition cross-sections.

\begin{figure}[h]
	\centering
	\begin{minipage}{0.4\textwidth}
		\centering
		\includegraphics[width=\linewidth]{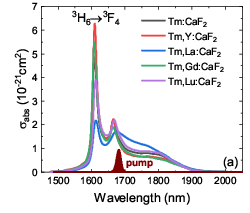}
	\end{minipage}
	\begin{minipage}{0.4\textwidth}
		\centering
		\includegraphics[width=\linewidth]{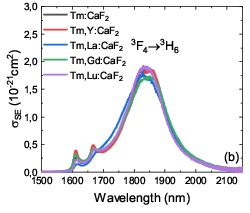}
	\end{minipage}	
	
	\vspace{0.5em}
	
	\begin{minipage}{0.4\textwidth}
		\centering
		\includegraphics[width=\linewidth]{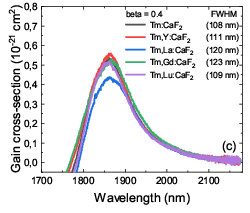}
	\end{minipage}
	\begin{minipage}{0.4\textwidth}
		\centering
		\includegraphics[width=\linewidth]{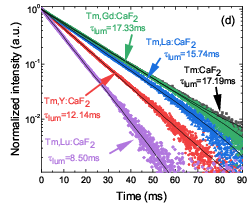}
	\end{minipage}
	
	\caption{Spectroscopy of Tm$^{3+}$ ions in CaF$_2$ crystals with and without buffers X = Y, La, Gd and Lu:  
		(a) absorption cross-sections, $\sigma_{\mathrm{abs}}$, for the $^3$H$_6 \rightarrow ^3$F$_4$ transition, the spectrum of the InGaAsP/InP pump diode is shown for comparison;  
		(b) stimulated-emission (SE) cross-sections, $\sigma_{\mathrm{SE}}$, for the $^3$F$_4 \rightarrow ^3$H$_6$ transition;  
		(c) gain spectral profiles, $\sigma_{\mathrm{g}} = \beta \sigma_{\mathrm{SE}} - (1 - \beta) \sigma_{\mathrm{abs}}$, for an inversion ratio $\beta = 0.4$;  
		(d) fluorescence decay curves, $\lambda_{\mathrm{exc}}$=1.62~\textmu m, $\lambda_{\mathrm{lum}}$=1.86~\textmu m, $\tau_{\mathrm{lum}}$ – luminescence lifetime.}
	\label{fig:spectroscopy}
\end{figure}

The absorption cross-sections $\sigma_{\mathrm{abs}}$ for the $^3$H$_6 \rightarrow ^3$F$_4$ transition of Tm$^{3+}$ ions in Tm,X:CaF$_2$ crystals used for resonant pumping are shown in Fig.~\ref{fig:spectroscopy}(a). They were calculated using the ion density $N_{\mathrm{Tm}} = 3.65 \times 10^{20}$ at/cm$^3$ (the segregation coefficient for Tm$^{3+}$, $K_{\mathrm{Tm}} \sim 1$). 

Depending on the nature of the buffer ion, the peak $\sigma_{\mathrm{abs}}$ at 1.61~µm varies significantly, ranging from $2.18 \times 10^{-21}$~cm$^2$ (for Tm,La:CaF$_2$) to $6.28 \times 10^{-21}$~cm$^2$ (for Tm,Y:CaF$_2$). For the latter crystal, the absorption bandwidth (full width at half maximum) is 16~nm. These absorption cross-sections exceed those for the $^3$H$_6\rightarrow ^3$H$_4$ transition at 0.8~µm (in the respective compositions), highlighting the potential of resonant pumping for boosting the pump absorption efficiency in Tm-lasers. The $\sigma_{\mathrm{abs}}$ values at the pump wavelength employed in the present work (1.68~µm) are listed in Tab.~\ref{tab:spectroscopic_parameters}.

The stimulated-emission (SE) cross-sections $\sigma_{\mathrm{SE}}$ for the $^3$F$_4 \rightarrow ^3$H$_6$ laser transition were calculated from the measured fluorescence spectra $I(\lambda)$ (corrected for the structured water vapor absorption) using the Füchtbauer-Ladenburg equation~\cite{Aull1982}:

\begin{equation}
	\sigma_{\mathrm{SE}}(\lambda)=\frac{\lambda^5}{8 \pi c\langle n \rangle ^2\tau_{rad}}\frac{I(\lambda)}{\int I(\lambda)\lambda d\lambda},
\end{equation}

where $\lambda$ is the emission wavelength, $\langle n \rangle \approx 1.42$ is the mean refractive index of the CaF$_2$ crystal~\cite{Malitson1963}, $c$ is the speed of light, and $\tau_{\mathrm{rad}} = 21.3$~ms is the radiative lifetime of the $^3$F$_4$ state~\cite{Loiko2021}. A very small effect of the crystal composition on this value was observed (within the error of the analysis). The SE cross-sections are given in Fig.~\ref{fig:spectroscopy}(b). Their spectral dependence is less sensitive to the nature of the buffer ion. Around 1.9~µm, $\sigma_{\mathrm{SE}}$ lies in the range of 1.0–1.2 $\times 10^{-21}$~cm$^2$ for all the studied crystals, see Tab.~\ref{tab:spectroscopic_parameters}.

According to the quasi-3-level operation scheme of Tm lasers with reabsorption, the gain cross-sections, $\sigma_{\mathrm{g}} = \beta \sigma_{\mathrm{SE}} - (1 - \beta) \sigma_{\mathrm{abs}}$, were calculated. They are compared for the studied Tm,X:CaF$_2$ crystals for a given inversion ratio $\beta = 0.4$ (where $\beta = N_2 / N_{\mathrm{Tm}}$, $N_2$ is the population of the upper laser level, $^3$F$_4$). The gain profiles are broad and almost structureless, and they extend beyond 2~µm, see Fig.~\ref{fig:spectroscopy}(c). The gain bandwidth (FWHM) was calculated, cf. Tab.~\ref{tab:spectroscopic_parameters}, ranging from 108~nm (for Tm:CaF$_2$) to 123~nm (Tm,Gd:CaF$_2$).

The luminescence decay curves from the $^3$F$_4$ manifold measured under resonant excitation are plotted in Fig.~\ref{fig:spectroscopy}(d). Finely powdered crystalline samples were used to reduce the effect of radiation trapping (reabsorption). The decay is close to single-exponential, suggesting a single type of Tm (or Tm–X) clusters (“quasi-centers”, as proposed by Kaminskii~\cite{Kaminskii2020}). The nature of the buffer has a significant influence on the upper laser level lifetime, which ranges from 8.50~ms (for Tm,Lu:CaF$_2$) to 17.33~ms (for Tm,Gd:CaF$_2$), see Tab.~\ref{tab:spectroscopic_parameters}.

From the point of view of spectroscopic properties, generally speaking, one should select a crystal combining:  
i) high absorption cross-sections (for efficient pumping),  
ii) the broadest gain profiles extending far beyond 2~µm (for broadband wavelength tuning and reduced impact of water vapor absorption), and  
iii) the longest luminescence lifetime (for low laser thresholds).  
The Tm,Gd:CaF$_2$ crystal appears as the most promising candidate. At the same time, these criteria can be conflicting, e.g., the strongest spectral broadening will inevitably lead to lower transition cross-sections.

\begin{table}[h]
	\centering
	\begin{tabular}{lllll}
		\hline
		Crystal &
		\begin{tabular}[c]{@{}l@{}}$\sigma_{\mathrm{abs}}$,\\ 10$^{-21}$ cm$^2$ \\ at 1.68 \textmu m\end{tabular} &
		\begin{tabular}[c]{@{}l@{}}$\sigma_{\mathrm{SE}}$,\\ 10$^{-21}$ cm$^2$ \\ at 1.9 \textmu m\end{tabular} &
		$\tau_{\mathrm{lum}}$, ms &
		\begin{tabular}[c]{@{}l@{}}$\Delta \lambda_g$, nm,\\ for $\beta$ = 0.4\end{tabular} \\
		\hline
		Tm:CaF$_{2}$    & 1.43          & 1.04          & 17.2          & 108          \\
		Tm,Y:CaF$_{2}$  & 1.29          & 1.10          & 12.1          & 111          \\
		Tm,La:CaF$_{2}$ & \textbf{1.69} & 1.06          & 15.7          & 120          \\
		Tm,Gd:CaF$_{2}$ & 1.31          & \textbf{1.18} & \textbf{17.3} & \textbf{123} \\
		Tm,Lu:CaF$_{2}$ & 1.67          & 1.01          & 8.5           & 109 \\
		\hline         
	\end{tabular}
	
	\caption{Key spectroscopic parameters of Tm$^{3+}$ ions in the Tm,X:CaF$_{2}$ crystals. $\sigma_{\mathrm{abs}}$ – absorption cross-section for in-band pumping,  
		$\sigma_{\mathrm{SE}}$ – stimulated-emission cross-section,  
		$\tau_{\mathrm{lum}}$ – luminescence lifetime,  
		$\Delta\lambda_{\mathrm{g}}$ – gain bandwidth (FWHM).}
	\label{tab:spectroscopic_parameters}
	
\end{table}

\section{Resonantly pumped laser operation}

\subsection{Laser set-up}

\begin{figure}[h]
	\centering
	\includegraphics[width=0.5\textwidth]{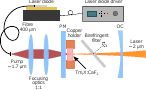}
	\caption{Layout of the wavelength tunable resonantly diode-pumped Thulium laser.}
	\label{fig:Scheme}
\end{figure}

The layout of the laser setup is depicted in Fig.~\ref{fig:Scheme}. For laser experiments, the crystals were polished to laser grade quality and left uncoated. They were mounted in a massive copper holder cooled by circulating water (15~$^\circ$C) using indium foil for better thermal contact.
A near hemispherical resonator was used, formed by a plane dichroic pump mirror coated for high reflectance (HR) at 1.9–2.15~\textmu m and high transmission (HT, $T > 92\%$) at 1.68~µm, and a set of curved output couplers (OC) with reflectivity at the laser wavelength $R_{OC}$ ranging between 92.4\% to 99.8\%, and a radius of curvature ($r_{OC}$) of -100~mm or -150~mm. The laser crystal was positioned 3~mm away from the pump mirror.
The pump source was a fiber-coupled (fiber core diameter: 400~\textmu m, N.A.~=0.22) InGaAsP/InP laser diode (QPC Lasers, model BrightLase Ultra-100 6217-0003) emitting up to 30~W of unpolarized radiation at a central wavelength of 1.68~\textmu m. Although the absorption cross-section of Tm$^{3+}$ in our crystals peaks near 1.61~\textmu m (Fig.~2a), no pump source was available at this wavelength. The chosen 1.68~\textmu m emission still lies within the broad $^3$H$_6 \rightarrow ^3$F$_4$ absorption band of Tm$^{3+}$ ions, ensuring efficient in-band pumping. With increasing driving current, the emission wavelength exhibited a red-shift of about 0.2~nm/A, and the measured laser linewidth (FWHM) was approximately 16~nm.
The output from the diode was collimated and focused into the crystal using a pair of achromatic doublets with a focal length of $f = 75$~mm each, resulting in a 1:1 reimaging ratio and a measured pump spot radius of $190 \pm 10$~\textmu m in the focus. The diode emission was characterized by the measured half divergence angle of $163 \pm 5$~mrad, and a beam quality factor $M^2$ of $58 \pm 4$ (measured behind the focusing optics).
The pump radiation was electronically modulated, resulting in 10~ms pulses at a repetition rate of 10~Hz (pulse pumping). The pumping was single pass. The measured small-signal pump absorption efficiency was in the range of 21\% to 30\%, depending on the buffer ion, in agreement with the absorption spectra, and on increasing the pump level, it further dropped by 8\% due to a combined effect of the ground-state bleaching and red-shift of the diode wavelength.

The mean output power of the laser radiation was measured using a Thorlabs PM100USB power meter equipped with an S401C sensor. The spectra of both the pump and laser radiation were recorded using a StellarNet RED-Wave spectrometer. The beam profile was captured with a Spiricon Pyrocam IV camera (pixel size: $80 \times 80$~µm$^2$).

The tuning of the laser wavelength was enabled by a birefringent filter (BRF) made of MgF$_2$, which is a positive uniaxial material. The BRF had a diameter of 20~mm and a thickness of 1~mm, with its optical axis lying in the plane of the filter. It was placed in the resonator at the Brewster angle, close to the laser crystal. 

At the wavelength around 2~\textmu m, the free spectral range (FSR) of the filter was 288--440~nm depending on the rotation angle. The filter was mounted in a motorized rotation holder (SmarAct, model SR-5714). During the wavelength tuning, the average output power and the spectrum of the laser radiation were simultaneously measured for different filter rotation angles with a step of 1.7~mrad.

\subsection{Free-running laser operation}

\begin{figure}[h]
	\centering
	\begin{minipage}{0.4\textwidth}
		\centering
		\includegraphics[width=\linewidth]{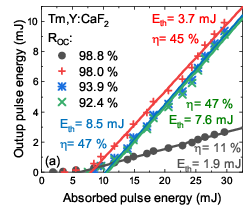}
	\end{minipage}
	\begin{minipage}{0.4\textwidth}
		\centering
		\includegraphics[width=\linewidth]{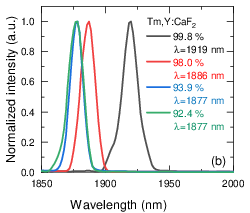}
	\end{minipage}

	\caption{Resonantly diode-pumped Tm,Y:CaF$_2$ laser operating in the free-running regime:  
		(a) power transfer curves, duty cycle: 1:10, $\eta$ – slope efficiency, $P_{th}$ – laser threshold (vs. absorbed pump pulse energy);  
		(b) typical spectra of laser emission (unpolarized output).}
	\label{fig:TmYCaF2}
\end{figure}

First, free-running laser operation was exploited (without the BRF in the cavity). The input–output characteristics for different output coupling rates for the best-performing crystal Tm,Y:CaF$_2$ are shown in Fig.~\ref{fig:TmYCaF2}(a). 

An output pulse energy of 9.4~mJ was achieved at an absorbed pump pulse energy of 29.6~mJ, corresponding to a slope efficiency $\eta$ of 47\% and a laser threshold $P_{th}$ of 8.5~mJ (for $R_{OC}$ = 93.9\%). On decreasing the output coupler reflectivity from 99.8\% to 92.4\%, the laser threshold increased from 1.9~mJ to 7.6~mJ.

The typical spectra of laser emission measured well above the laser threshold are presented in Fig.~\ref{fig:TmYCaF2}(b), revealing a blue-shift of the laser wavelength for higher output coupling rates, from 1919~nm ($R_{OC}$ = 99.8\%) to 1877~nm ($R_{OC}$ = 92.4\%). This behavior is typical for quasi-3-level Tm lasers with reabsorption and aligns well with the gain profiles.

The measured duration of laser pulses increased with the pump level in the range of 0.2 to 11~ms.

\begin{figure}[h]
	\centering
	\begin{minipage}{0.4\textwidth}
		\centering
		\includegraphics[width=\linewidth]{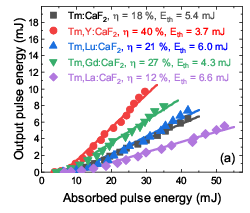}
	\end{minipage}
	\begin{minipage}{0.4\textwidth}
		\centering
		\includegraphics[width=\linewidth]{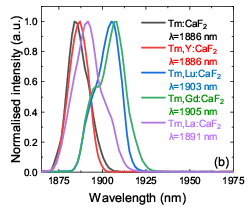}
	\end{minipage}

	\caption{Effect of the rare-earth buffer (X = Y, La, Gd and Lu) incorporation on the output performance of resonantly diode-pumped Tm,X:CaF$_2$ lasers:  
		(a) input-output dependences, $\eta$ – slope efficiency, $P_{th}$ – laser threshold (vs. absorbed pump pulse energy);  
		(b) laser emission spectra. Output coupler reflectivity: $R_{OC}$ = 98\%, free-running regime.
	}
	\label{fig:TmXCaF2}
\end{figure}

Fig.~\ref{fig:TmXCaF2} compares the output characteristics (power transfer curves and typical laser spectra) of all five tested laser crystals for a given output coupler reflectivity of 98.0\%. Note that the different pump absorption efficiencies determine the range of the absorbed pump pulse energies. The maximum incident pump pulse energy was the same: 221~mJ. There is a significant variation in laser efficiency with the crystal composition, which is mostly associated with the crystal quality. The best performance was achieved for the Tm,Y:CaF$_2$ crystal.

The typical beam profiles in the far-field for the Tm,X:CaF$_2$ lasers are compared in Fig.~6. The lasers operated on the fundamental transverse mode. Based on these profiles, the beam quality factor was estimated as M$^2<1.6$ for all tested samples.

\begin{figure}[h]
	\centering
	\includegraphics[width=0.7\textwidth]{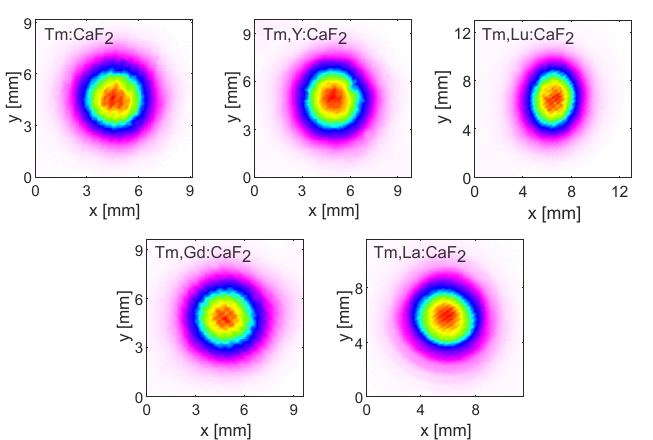}
	\caption{Far-field beam profiles of the in-band diode-pumped Tm,X:CaF$_2$ lasers (taken at $\sim$40 cm from the OC).}
	\label{fig:profiles}
\end{figure}

\subsection{Wavelength tuning}
\begin{figure}[h]
	\centering
	\begin{minipage}{0.4\textwidth}
		\centering
		\includegraphics[width=\linewidth]{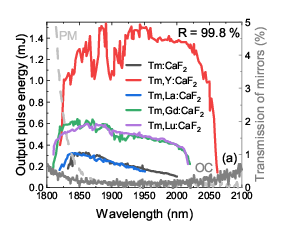}
	\end{minipage}
	\begin{minipage}{0.4\textwidth}
		\centering
		\includegraphics[width=\linewidth]{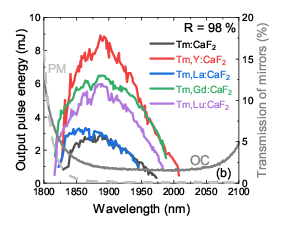}
	\end{minipage}	
	
	\vspace{0.5em}
	
	\begin{minipage}{0.5\textwidth}
		\centering
		\includegraphics[width=\linewidth]{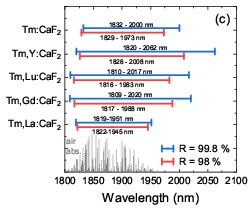}
	\end{minipage}
	
	\caption{Results of the laser tuning experiments for resonantly diode-pumped Tm,X:CaF$_2$ lasers (X = no buffer, Y, La, Gd and Lu): (a,b) Output coupler reflectivity: (a) $R_{OC}$ = 99.8\%, (b) $R_{OC}$ = 98\%; (c) an overview of the obtained tuning ranges at the zero power; grey lines – water vapor absorption (HITRAN database, https://hitran.org/).
	}
	\label{fig:tuning}
\end{figure}

The wavelength tuning experiment was performed for two output couplers with reflectivities $R_{OC}$ of 99.8\% and 98.0\%, under identical pumping conditions (incident pump pulse energy of 221 mJ, pump pulse length of 10 ms, repetition rate of 10 Hz). The tuning curves are given in Fig.~\ref{fig:tuning}(a,b). They are relatively flat and broad, with the wavelength being continuously tuned within the corresponding ranges. The broadest tuning range, accompanied by the highest output energy, was achieved for the Tm,Y:CaF$_2$ crystal in both cases. For the higher-reflectivity mirror, the tuning range was 1820--2062 nm (242 nm at the zero-power level), and for the lower-reflectivity mirror, it was narrower, measuring 1826--2008 nm (182 nm). Figure~7(c) summarizes the tuning ranges for all the studied crystals.

\subsection{Discussion}
Let us discuss the criteria for the choice of “buffer” ions aiming the material engineering of highly performant Tm,X:CaF$_2$ laser gain media for broadly tunable 2~\textmu m lasers. While Tm,Gd:CaF$_2$ fulfills most of the spectroscopic criteria, the realized laser output is governed by the passive losses. In our set, Tm,Y:CaF$_2$ exhibited a lower threshold and higher slope efficiency (Fig.~5a), which we attribute to its lower passive loss (higher optical quality). The crystals used here were primarily fabricated for spectroscopic studies, i.e., not grown in laser-grade quality; this could have penalized Tm,Gd:CaF$_2$ more strongly. Nevertheless, Tm,Gd:CaF$_2$ delivered the second-best laser performance and tuning range in our set, suggesting that with improved growth its superior spectroscopic properties could translate into higher slope efficiency and broader tuning range. Furthermore, the difference between the ionic radii of the Tm$^{3+}$ (1.13 \AA), X$^{3+}$ (from 1.11 \AA\ for Lu$^{3+}$ to 1.32 \AA\ for La$^{3+}$) and Ca$^{2+}$ (1.26 \AA, all the values are given for VIII-fold fluorine coordination \cite{Kaminskij1990}) may be the main factor explaining the variations in the optical quality of grown crystals.

Previous structural studies on rare-earth ion clusters in CaF$_2$ – RF$_3$ fluorite-type solid-solutions \cite{Sulyanova2022,Sorokin2014,Ivanov-Shits1990,Sobolev2003} revealed their two main geometries of different sizes: tetrahedral (small) R$_4$F$_{26}$ ones and octahedrocubic (large) Ca$_8$R$_6$F$_{69}$ or Ca$_9$R$_5$F$_{69}$ ones, which substitute for the Ca$_4$F$_{23}$$^{15-}$ and Ca$_{14}$F$_{64}$$^{36-}$ building blocks of the host structure, respectively (the so-called block isomorphism). Their formation is evidenced by the presence of F$^-_{\mathrm{in}}$, respectively, 32$f$ sites and 48$i$ sites. The local charge compensation is maintained by mobile carriers F$^-_{\mathrm{mob}}$ at the periphery of these defect regions. Depending on the nature of the rare-earth ion, these solid-solutions can be divided into three groups: i) those with only tetrahedral clusters (R = La to Tb), ii) those with only octahedrocubic ones (R = Er to Lu), and iii) those in which both geometries can coexist (R = Y, Dy, Ho).

Physically, strong (or weak) effect of the introduction of “buffer” ions into the Tm$^{3+}$-doped CaF$_2$ crystal can be explained as follows: for crystals singly doped with Tm$^{3+}$, only octahedrocubic clusters exist (with an octahedral arrangement of R$^{3+}$ and Ca$^{2+}$ in the first coordination sphere and cubic arrangement of Ca$^{2+}$ in the second coordination sphere). This well agrees with the nearly single-exponential decay of Tm$^{3+}$ ions in heavily doped (a few at.\%) CaF$_2$ crystals suggesting the existence of a single “quasi-center”. By codoping the crystal with “buffer” cations belonging to the same group (e.g., Lu$^{3+}$), one still expects the Ca$_8$R$_6$F$_{69}$ or Ca$_9$R$_5$F$_{69}$ clusters. Indeed, the effect of Lu$^{3+}$ on the variation of the absorption and emission profiles is weak. In contrast, by codoping with cations that favor the formation of only tetrahedral clusters (i.e., La$^{3+}$ or Gd$^{3+}$), a much stronger disorder is introduced which is indeed reflected in the spectral profiles. The case of Tm$^{3+}$,Y$^{3+}$:CaF$_2$ is exceptional as two types of clusters are expected to coexist (note that for this crystal, the strongest deviation from Tm:CaF$_2$ as can be seen in Fig.~\ref{fig:spectroscopy}(a,b)).

Further optimization of the material composition should focus on finding the best Tm/X codoping ratio for a given Tm concentration. Note that for in-band pumping which does not rely on cross-relaxation, efficient laser operation even at low Tm doping levels is possible as long as the combination of the doping concentration and the sample length (e.g., in the “rod” geometry) provides sufficient pump absorption efficiency. The rare-earth doping into CaF$_2$ (accounting for both laser-active and “buffer” ions) rapidly decreases the thermal conductivity of the crystal leading to a higher probability of thermal fracture which is already a serious issue for an overall doping level of 5 at.\% \cite{Liu2025, Druon2011}. It seems that this decline in thermal properties is nearly the same for all the rare-earth ions, as demonstrated by Pavel Popov in his systematic studies \cite{Popov2008,Popov2017}. Therefore, it is desirable to keep the Tm$^{3+}$ doping level low (about 0.3 to 1 at.\%) and optimize the Tm/X codoping ratio (1:2 to 1:5) to induce a noticeable variation in the crystal-field within rare-earth clusters which are readily formed at an overall doping level exceeding 0.5 at.\% (i.e., the targeted range for the “buffer” ions is 0.6 to 5 at.\%).

\section{Conclusion}
To conclude, we report on the first comparative study of five Tm$^{3+}$-doped calcium fluoride crystals singly doped and codoped with optically inactive “buffer” cations (Y$^{3+}$, La$^{3+}$, Gd$^{3+}$ and Lu$^{3+}$), regarding their widely tunable laser operation in the eye-safe spectral range of 2 \textmu m. As the main parameter for assessing crystal performance, we chose the wavelength tuning range, which in fact combines the contributions from spectroscopic behavior (e.g., via the accessible gain bandwidth and the upper-laser level lifetime) and the optical quality of the crystals (via the total intracavity losses). We also employed the in-band pumping scheme at 1.68 \textmu m which does not rely on the cross-relaxation mechanism and consequently allows one to operate at lower Tm$^{3+}$ doping levels (1.5 at.\%, in the present work) accessing better thermal properties of the crystals. 

We identify the Tm,Y:CaF$_2$ crystal as the most promising candidate for widely tunable continuous-wave and potentially femtosecond mode-locked lasers, as in the free-running regime, it enables a remarkable tuning range of 242 nm (1820–2062 nm), extending far beyond 2 \textmu m even despite the inherently weak crystal-field strengths in fluoride matrices. This crystal also demonstrates a laser slope efficiency up to 47\%, which could be further improved by optimizing the pump-laser mode overlap in the resonator, as well as reducing the Tm$^{3+}$ doping level to suppress ETU from the upper laser manifold. We suggest an explanation for this remarkable performance, by analyzing the variety of rare-earth cluster types. 

We believe that true continuous-wave laser operation in the multi-Watt regime is possible when applying proper cooling of laser crystals (e.g., in the rod geometry) and reducing the overall rare-earth (Tm$^{3+}$ + X$^{3+}$) doping to a few at.\%. Another way to extend the tuning range far beyond 2 \textmu m would be to explore in-band pumping of triply doped (Tm$^{3+}$, Ho$^{3+}$, X$^{3+}$) crystals.

\begin{backmatter}

\bmsection{Acknowledgement}
Preliminary results were presented at CLEO/Europe-EQEC 2025 \cite{Popelova2025b}.

\bmsection{Funding}
This work has been funded by a grant from the Programme Johannes Amos Comenius under the Ministry of Education, Youth and Sports of the Czech Republic from the project LASCIMAT, project No. CZ.02.01.01/00/23\_020/0008525. As set out in the Legal Act, beneficiaries must ensure that the open access to the published version or the final peer-reviewed manuscript accepted for publication is provided immediately after the date of publication via a trusted repository under the latest available version of the Creative Commons Attribution International Public Licence (CC BY) or a licence with equivalent rights. This work has been also funded by French Agence Nationale de la Recherche (ANR-22-CE08-0031, FLAMIR) and Région Normandie, France (Contrat de plan État-Région (CPER)).

\bmsection{Disclosures}
The authors declare no conflicts of interest.

\bmsection{Data Availability Statement}
Data underlying the results presented in this paper are available in dataset "Tm,X:CaF$_2$ lasers", Ref. \cite{Popelova2025a}.

\end{backmatter}

\label{sec:refs}

\bibliography{References_final}

\end{document}